\documentclass[twocolumn,twocolappendix]{aastex631}

\usepackage{lipsum}

\newcommand{\tarcsec}{\text{arcsec}}

\received{April 24, 2025}
\revised{May 28, 2025}
\accepted{May 30, 2025}

\submitjournal{ApJL}

\begin{document}

% Hmm, alternate titles....
% Maybe:

% The Intracluster Light of Abell 3667: Unveiling an Optical Bridge with DECam
% That's true to the project, but I want to highlight the LSST connection

% The Intracluster Light of Abell 3667: Unveiling an Optical Bridge in LSST Precursor Data from DECam
% That hammers LSST and DECam, but is getting long?

% The Intracluster Light of Abell 3667: Unveiling an Optical Bridge in LSST Precursor Data
% Doesn't highlight the instrument, but not bad

\title{The Intracluster Light of Abell 3667: Unveiling an Optical Bridge in LSST Precursor Data}

\author[0000-0003-2314-5336]{Anthony M. Englert}
\affiliation{Department of Physics, Brown University, Providence, RI 02912, USA}

\author[0000-0003-0751-7312]{Ian Dell'Antonio}
\affiliation{Department of Physics, Brown University, Providence, RI 02912, USA}

\author[0000-0001-7847-0393]{Mireia Montes}
\affiliation{Institute of Space Sciences (ICE, CSIC), Campus UAB, Carrer de Can Magrans, s/n, 08193 Barcelona, Spain}



\begin{abstract}

Intracluster light, the diffuse glow of stars stripped from galaxies during a cluster's formation, is an established tracer of a cluster's dynamical history. The upcoming Vera C. Rubin Observatory's Legacy Survey of Space and Time (LSST) is set to revolutionize studies of intracluster light by imaging the entire southern sky down to a limiting surface brightness $\mu \gtrsim 30\text{mag}/\tarcsec^2$ by year ten. In this letter, we create a precursor LSST dataset (reaching the equivalent of year eight depth) using DECam observations of Abell 3667 and study its intracluster light. We have discovered a low surface brightness ($ \mu \gtrsim 26\text{mag}/\tarcsec^2 $) optical bridge extending over $\sim 400\text{ kpc}$ which connects the two brightest galaxies (BCG1 and BCG2) in the cluster; the color and surface brightness of the bridge is consistent with formation via a major merger. The inner regions of BCG1 ($r < 200\text{ kpc}$) and BCG2 ($r < 50\text{ kpc}$) are consistent with formation via gradual stripping of satellite galaxies, but BCG2's outer profile appears disrupted by a recent merger.
%We find that the intracluster light near BCG1 ($r < 200\text{ kpc}$) is consistent with formation via the gradual stripping of satellite galaxies. The inner-regions of the BCG2 ($r < 50\text{ kpc}$) are consistent with gradual stripping as well but, beyond that, there is evidence that the profile has been disrupted by a recent merger. 
We hypothesize that the bridge is a relic of a recent first-pass between the two brightest galaxies and is composed of stars being stripped from BCG2. Future studies of intracluster light with LSST will discover new features such as the bridge in local clusters while enabling detailed studies of the stellar populations of these features with its six photometric bands.

% need to add stuff here; 

\end{abstract}



\section{Introduction} \label{sec:intro}

Galaxy clusters are among the largest gravitationally bound objects in the universe and span a variety of dynamical states, ranging from recent mergers to fully virialized systems \citep{kravtsov_formation_2012}. 
Actively merging clusters in particular can be used to constrain the nature of dark matter \citep{harvey_nongravitational_2015,clowe_direct_2006} and probe cosmology \citep{allen_cosmological_2011}.
They show abundant evidence of their history through diffuse X-ray and radio emissions arising from interactions within the intracluster medium \citep{sarazin_x-ray_1986,van_weeren_diffuse_2019}.
These emissions can be used to infer the dynamical history based on the presence of radio relics \citep{lee_radio_2024,finner_weak-lensing_2025} or X-ray features such as shock fronts \citep{sarazin_x-ray_1986}.
In optical studies, a complimentary tracer of a cluster's dynamical history is the intracluster light (ICL).

ICL is the combined emission of individual stars which have been stripped from cluster members by tidal forces throughout the cluster's formation \citep{contini_origin_2021,montes_faint_2022}. It is a feature in the low surface brightness (LSB) regime, with characteristic surface brightnesses fainter than $\mu_V \sim 26\text{ mag}/\tarcsec^2$, and is often observed by studying the surface brightness profile of a brightest cluster galaxy galaxy (BCG) out to large distances ($\gtrsim 100\text{ kpc}$) from the core. 
The ICL is an established tracer for the distribution of dark matter across a cluster \citep{diego_exploring_2023,montes_intracluster_2019,sampaio-santos_is_2021,cha_high-caliber_2025,yoo_comparison_2022,yoo_spatial_2024,alonsoasensio_intracluster_2020} and, by studying the the surface brightness profile and color profiles of the combined BCG+ICL system, the dynamical history of a cluster can be inferred.

The ICL can form through various ``formation channels''; the most relevant ones for this letter are the stellar stripping and merger scenarios. In the stellar stripping scenario, ICL is formed from the stripping of stars from satellite galaxies near the BCG. This occurs gradually as the cluster evolves, so older and lower-metallicity stars stripped early in the cluster's history will be closer to the core of the BCG, creating a negative color gradient (from red to blue) across the profile of the BCG+ICL system \citep{contini_different_2018,demaio_origin_2015,demaio_lost_2018, montes_intracluster_2014, montes_buildup_2021}. In the merger scenario, ICL is formed from the rapid stripping of stars from galaxies during a merger; this mixes the ages and metallicities of stars across the ICL, producing a flat color-profile \citep{contini_origin_2021,montes_intracluster_2018}. The distinct impact of these formation channels can be used to infer the presence of mergers in a cluster's dynamical history \citep{mihos_burrell_2016,iodice_intracluster_2017,demaio_lost_2018, montes_new_2022}.

Abell 3667 (A3667) is an actively merging cluster featuring complex X-ray emissions and prominent radio relics \citep{gasperin_meerkat_2022,storm_xmm-newton_2018,carretti_detection_2013,owers_substructure_2009,lovisari_metallicity_2009}. Currently, the leading hypothesis is that A3667 is the product of an offset intermediate mass merger, which can simultaneously explain the origin of the radio relics and the structure of the X-ray emissions \citep{gasperin_meerkat_2022,omiya_indications_2024}. Studying the ICL of A3667 offers a compelling alternate method of confirming this hypothesis. However, A3667's low redshift ($z=0.0556$) constrains this analysis to deep-wide ground-based imaging where the large field of view can capture the ICL.

Generally, ICL studies originate from either space-bound observations of high redshift ($z > 0.1$) clusters \citep{montes_intracluster_2014,burke_coevolution_2015,morishita_characterizing_2017,jimenez-teja_unveiling_2018,montes_intracluster_2018,gonzalez_discovery_2021,montes_new_2022,martis_modeling_2024,ellien_euclid_2025} or from ground-based observations of local clusters with deep-wide field imaging \citep{mihos_diffuse_2005,gonzalez_intracluster_2005,iodice_intracluster_2017,zhang_dark_2019,kluge_structure_2020,montes_buildup_2021,jimenez-teja_deep_2025}. The Rubin Observatory's Legacy Survey of Space and Time (LSST) is set to revolutionize ICL studies of local clusters by observing the entire southern sky down to a limiting surface brightness of at least $\mu \simeq 30.3 \text{ mag}/\tarcsec^2$ by year 10 \citep{brough_preparing_2024,bianco_optimization_2022}. Since the LSST Science Pipelines (LSP) are built to process data from multiple observatories \citep{bosch_hyper_2018,bosch_overview_2019}, precursor datasets of similar depth can be built using the latest algorithms intended for LSST \citep{aihara_hyper_2018,fu_lovoccs_2022,luo_merian_2024}.

In this letter, we present a precursor LSST dataset for A3667 and study its ICL; the remainder of the letter is structured as follows. In Section \ref{sec:dataproc}, we summarize our observations, implementation of the LSP, and post-LSP processing steps to create coadds suitable for LSB science. In Section \ref{sec:icl}, we characterize the ICL of A3667, and in Section \ref{sec:conc}, we discuss how this is related to the cluster's dynamical history. Throughout, we use the AB magnitude system \citep{oke_secondary_1983} and assume a concordance $\Lambda$CDM cosmology with $H_0 = 70 \text{ km} \text{ s}^{-1} \text{ Mpc}$, $\Omega_m = 0.3$, and $\Omega_{\Lambda} = 0.7$. This produces a distance-scale of $1.08 \text{ kpc}/\tarcsec$ at $z=0.0556$. 

\section{Data and Processing} \label{sec:dataproc}

% manually shift this to center it
\begin{figure*}[htbp!]
    \centering
    \includegraphics[width=0.96\textwidth]{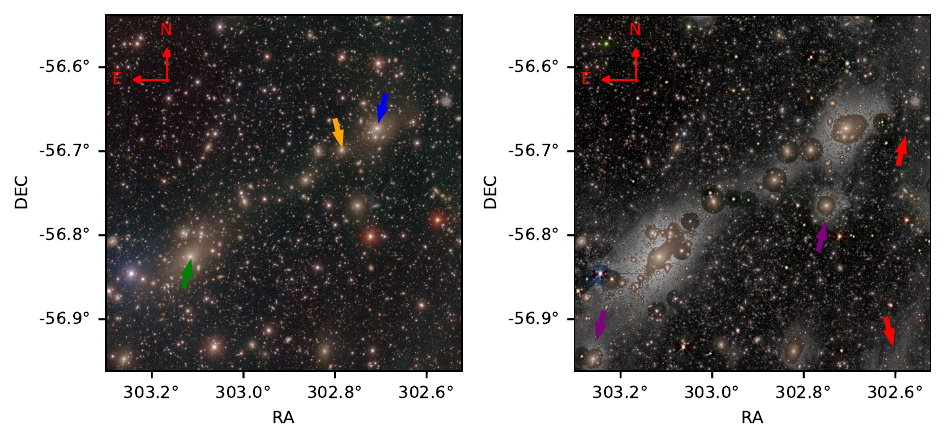}
    \caption{\textit{Left}: A $grz$ cutout of our sky-corrected coadd centered on the midpoint between BCG1 (green) and BCG2 (blue). An intermediate galaxy whose profile we study, LEDA 64218, is also labeled (yellow). \textit{Right}: A cutout of the same field from the sky-corrected and star-subtracted coadd with masked pixels (detections and stars) rendered in $grz$. The pixels rendered in grayscale show the $r$-band coadd with an aggressive stretch to showcase LSB features. Note the presence of cirrus clouds (red) and two cluster members which are enveloped by ICL (purple), { LEDA 64290 in the southeast and LEDA 95791 in the west}.}
    \label{fig:merged}
\end{figure*}

\subsection{Observations \& Data Processing}

%\begin{figure}
%    \includegraphics[width=0.48\textwidth]{figures/mw_profile_model_g_test.png}
%    \caption{The radial profile of the brightest star in the field in $g$-band (measured in calibrated flux units with a zero-point of $27$). The power-law model of the wings is appended to the Moffat core per \citep{jacobs_extended_2019} and the profile is fit following the algorithm outlined in Section \ref{sec:dataproc}.}
%    \label{fig:psf_fit}
%\end{figure}

%TODO should I talk more about why we chose DECam for our stack?
We used observations taken by the Dark Energy Camera (DECam) on the 4-m Blanco Telescope at the Cerro Tololo Inter-American Observatory (CTIO) \citep{flaugher_dark_2015}. We queried all public raw exposures taken in the $ugriz$ bands from the NOIRLab Astro Data Archive within $2^\circ$ of the X-ray luminosity peak based on the Meta Catalog of X-Ray Clusters \citep{piffaretti_mcxc_2011}. This query was wide enough to include stray pointings whose detectors overlap with the virial radius of the cluster ($\lesssim 1\text{ Mpc}$ of the X-ray peak), maximizing our depth in this region. A table of the proposal-IDs, PIs, and exposure times are provided in Appendix \ref{app:prop}.

%TODO what is the best way of formatting these little commands/code-bits?
Raw exposures were processed using a modified version of the \texttt{run\_steps} script built by the Local Volume Complete Cluster Survey (LoVoCCS) \citep{fu_lovoccs_2022}, which has been adapted to \texttt{v26\_0\_0} of the LSP \citep{englert_rcw,englert_aas_245}. Our implementation of the Data Release Pipeline for DECam (DRP)\footnote{\url{https://github.com/lsst/drp_pipe}} includes two additional features: a correction for the brighter-fatter effect \citep{coulton_exploring_2018,broughton_mitigation_2024} and an implementation of the \texttt{skycorr} algorithm originally built for the Hyper Suprime Cam (HSC) on the Subaru Telescope \citep{aihara_second_2019,aihara_third_2022}. Monthly bias and flat frames were built with the LSP's calibration-products pipeline\footnote{\url{https://github.com/lsst/cp_pipe}} and span the past decade of DECam's observations starting from the Dark Energy Survey \citep{collaboration_dark_2016}. These have been uploaded to Rubin Observatory's US Data Facility and are available upon request. A discussion of our photometric calibration has been summarized in Appendix \ref{app:depth}.

Background modeling has remained a challenge for ground-based deep-wide observations due to difficulties in separating extended LSB features (such as galactic cirri or ICL) from the true background in calibrated exposures \citep{li_reaching_2022,roman_galactic_2020}. To address this, \texttt{skycorr} removes the background by masking detected sources, subtracting a large scale gradient, a fitted skyframe, and lastly subtracting a small-scale local background (\texttt{bgmodel2}). The detection and background subtraction steps are iterated until the background converges, then the resulting ``sky-corrected" exposures are stacked to produce a coadd (Fig. \ref{fig:merged}). Like previous studies, we found that \texttt{bgmodel2} caused over-subtraction around the largest LSB features \citep{li_reaching_2022}, and disabled it during processing. This led to artifacts in the $i$-band and, to a lesser extent, the $z$-band due to their more complex sky backgrounds. Therefore, we limited our analysis of the ICL to the $g$ and $r$ band, while relying on $grz$-images for detection and manual inspection for features. { A detailed discussion of our implementation of \texttt{skycorr} is provided in Appendix \ref{app:sky}}.

{ 

\subsection{LSB Considerations}

{LSB science benefits greatly from dedicated observing strategies designed to minimize the background, including strategies such as creating sky-flats \citep{borlaff_missing_2019} or specialized dither patterns \citep{trujillo_beyond_2016}. Since our study uses archival data, we cannot apply these specialized observing strategies; we did, however, carry out extensive quality checks on each frame. Each exposure was manually inspected for defects and tight constraints were placed on the data that was stacked, including cuts on the seeing \citep[see ][]{fu_lovoccs_2022}. Additionally, we stacked exposures taken almost exclusively during dark-time; for the $< 7\%$ of frames taken during gray time, the moon was $>30^{\circ}$ from the target. Although no dedicated observing strategy was applied, previous proposals used dither-patterns sufficient to cover DECam's $\sim1\arcmin$ chip-gaps.}

The use of dome-flats can contribute to large- and small-scale variations in the background due to an uneven illumination \citep[e.g., ][]{2002ApJ...575..779F,borlaff_missing_2019,montes_buildup_2021}; both of these sources of uncertainty are implicitly included in our limiting surface brightness (Sec. \ref{sec:icl}). The monthly flats were manually inspected to verify that they were free of defects. We also executed a test to verify their stability: each pixel was normalized according to the mean flux across its detector, then the ratio of normalized-flats from adjacent months was computed. Assuming that the flats are stable, their ratios should have a mean ($\overline{r}$) near unity with a small standard deviation ($\sigma_r$). We found that, per-detector, successive flats were consistent to within $|1 - \overline{r}| \lesssim 10^{-5}$ with standard deviations $\sigma_r \lesssim 10^{-3}$. The quality of individual flats were verified similarly: per-detector, the normalized flats have standard deviations $\lesssim 10^{-2}$, i.e. they are uniform down to $\lesssim 1\%$ of the mean flux across a detector.}

\subsection{Star Subtraction}

Our implementation of \texttt{skycorr} preserves even undesirable LSB features including the wings of the extended point spread function (PSF), which overlap significantly with the ICL and will bias any measurements if left unaddressed. To remove the wings, we built a model of the extended PSF based on the brightest star in our coadd. After masking background sources and regions near the star with prominent LSB features, we fit a Moffat profile \citep{moffat_theoretical_1969} to the core ($r < 200 \text{px}$) and an exponential profile to the wings independently. These were spliced together according to the procedure outlined in \citet{infante-sainz_sloan_2020}.

To find the scaling of the PSF for a given star ($a$), we computed each star's radial profile ($f_*(r)$) and fitted it to the extended PSF ($f(r)$) by minimizing the chi-squared

\begin{equation}
    \chi^2 = \sum_i \left( f_*(r_i) - a f(r_i) \right)^2  \; .
\end{equation}

By fitting to a radial profile, rather than running a per-pixel optimization, our fit is less likely to be contaminated by background sources obscured by the extended PSF \citep{montes_buildup_2021}. Moreover, using SExtractor's segmentation image \citep{bertin_sextractor_1996}, we masked background sources, fitted and subtracted the model PSF, and iterated this until convergence. We found that two iterations was sufficient to converge on a reasonable profile for each star. 
% I agree with MC that this figure may not be necessary... especially since a similar process has been done before by MM

The PSF was modeled for each band individually; fitting and subtraction was carried out for all stars brighter than 14th magnitude in $g$ centered on their Gaia coordinates \citep{gaia_collaboration_gaia_2023}. This model does not account for variations in the PSF across the focal plane of each exposure. Therefore, we limited our subtraction and overall analysis to the central $0.88^{\circ} \times 0.88^{\circ}$ ($ 12\,000\text{px} \times 12\,000\text{px}$) region of the coadd, where the PSF is approximately symmetrical and shows little variation.

\subsection{Masking}

LSB science requires aggressive masking to avoid contamination from background and foreground sources that can obscure the extended features of interest. To mask these contaminants, we exported the detection and bad-pixel masks from the LSP. Additionally, to cover the diffraction spikes and other residuals not modeled by our radial PSF model, we masked all pixels within $\sim 39\arcsec$ ($150\text{px}$) of the core of each bright star.

Studying the profiles of BCGs and other galaxies in the field requires more care since contaminants will blend with and bias the measured profile. We used SExtractor's segmentation image to mask objects other than the target galaxy, fitted and subtracted a S\'ersic to the target, and iteratively masked/subtracted until the mask converged; we found that two iterations was sufficient. Each mask was inspected and any sources this algorithm missed were manually masked.

\section{ICL of A3667} \label{sec:icl}

\subsection{LSB Features}

\begin{figure*}[htbp!]
    \centering
    \includegraphics[width=0.96\textwidth]{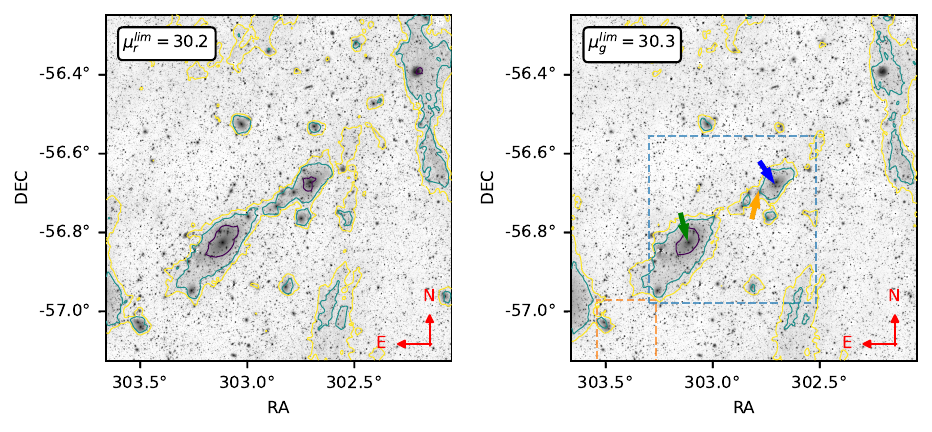}
    \caption{LSB contours of the central region of our coadds in $g$-band (left) and $r$-band (right) drawn at surface brightnesses of $26$, $28$, and $30 \text{mag}/\tarcsec^2$. These are drawn over the corresponding inverted star-subtracted coadd for each band. The eastern LSB feature is the extended PSF of a bright star outside the central region; the northern-most, northwestern, and southernmost $30 \text{mag}/\tarcsec^2$ contours are cirrus (the last two are also highlighted in Fig. \ref{fig:merged}). { BCG1 (green), BCG2 (blue), and LEDA 64218 (orange) are labeled in the right panel. Additionally, the blue square shows the field of view from Fig. \ref{fig:merged} and the orange rectangle shows the region displayed in Fig. \ref{fig:detectors}.}}
    \label{fig:lsb}
\end{figure*}

\begin{figure}[htbp!]
    \includegraphics[width=0.48\textwidth]{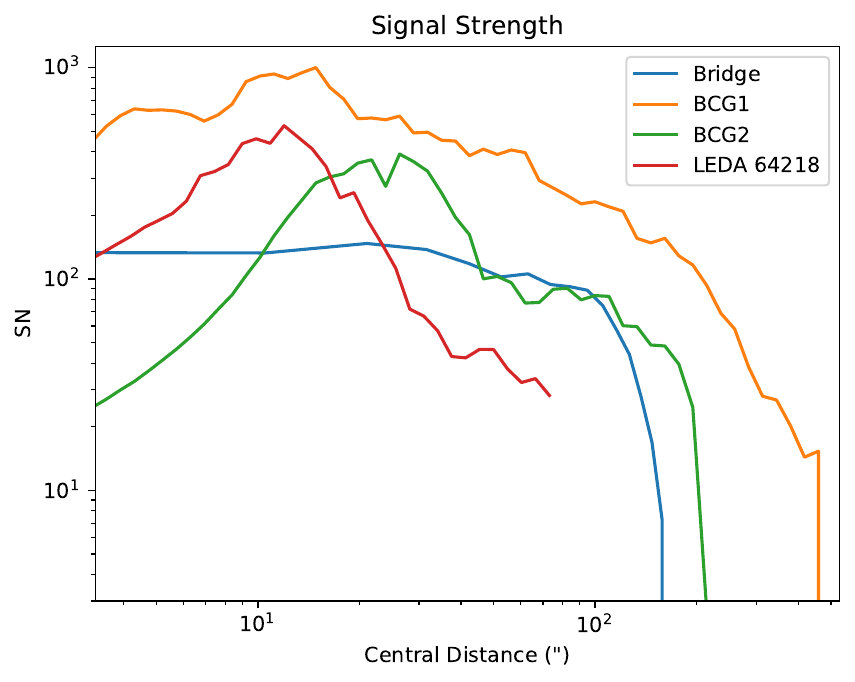}
    \caption{The signal-to-noise (SN) of the ICL-signal as a function of central distance for different features. The signal for LEDA 64218 is truncated since, beyond $\sim 100\arcsec$, the outer profile blends into the bridge. { Due to the small number of pixels used to fit isophotes near the core of BCG1, BCG2, and LEDA 64218, the SN-profile fluctuates and can have a peak offset from center of the object.}}
    \label{fig:sn}
\end{figure}

Cutouts of our coadds are presented in Fig. \ref{fig:merged}. The BCGs are connected by an optical-bridge and feature several prominent plumes connecting them to other cluster members. Galactic cirrus is also present and can be distinguished from the ICL by its morphology, color, and distance from the BCGs. 

To study the observed surface brightness of these features, we created LSB contours in the $g$ and $r$ bands. These were created from the masked and star-subtracted coadds, which were binned into $16\text{px}\times16\text{px}$ super-pixels and smoothed with a $\sigma=3$ super-pixel Gaussian filter. The limiting surface brightness was computed from the sky-corrected, but not star-subtracted, coadd following the procedure outlined in \citet{roman_galactic_2020} and is quoted for $3\sigma$ detections in $10\arcsec \times 10\arcsec$ squares (Fig. \ref{fig:lsb}). Since they are not representative of background pixels, the detected sources, ICL, and galactic cirri were masked during this calculation.

The LSB contours are shown in Fig. \ref{fig:lsb}. The optical bridge connecting the BCGs and their corresponding plumes extend out to the limiting surface brightness $\sim 30\text{mag}/\tarcsec^2$. The cirrus clouds extend well beyond the cutout showcased in Fig. \ref{fig:merged} and pose a risk of overlapping with the ICL, contaminating the signal. To verify that the signal was not contaminated, we computed the observed $g-r$ color enclosed by the $30\text{mag}/\tarcsec^2$ contours surrounding BCG1 and the southern cirrus cloud (Fig. \ref{fig:merged}) using the method outlined by \citet{roman_galactic_2020}. The color of the cirrus, $g-r \simeq 0.36 \pm 0.04$, is consistent with previous studies \citep{roman_galactic_2020,smirnov_prospects_2023,zhang_joint_2023} and distinct from the color of the ICL, $g-r \simeq 0.25 \pm 0.04$.

\subsection{Morphology}

The LSB features differ significantly from the otherwise elliptical morphology of cluster members. BCG1 features an extended plume to the southeast enclosing another member (LEDA 64290). The core of BCG2 is enclosed by shells and features two diffuse plumes ($\sim 30\text{ mag}/\tarcsec^2$) to the northeast and south.

% NEED MORE HERE?
The bridge itself appears to have a rectangular symmetry along a line connecting the cores of BCG1 and BCG2. Several cluster members overlap with the optical bridge and their extended profiles may contaminate the bridge's profile. To prevent this from biasing our analysis, we have masked all pixels within $\sim 53\arcsec$ ($200\text{px}$) of each overlapping cluster member (see right panel Fig.~\ref{fig:merged}).

\subsection{Surface Brightness \& Color Profiles}

\begin{figure*}[htbp!]
    \centering
    \includegraphics[width=0.98\textwidth]{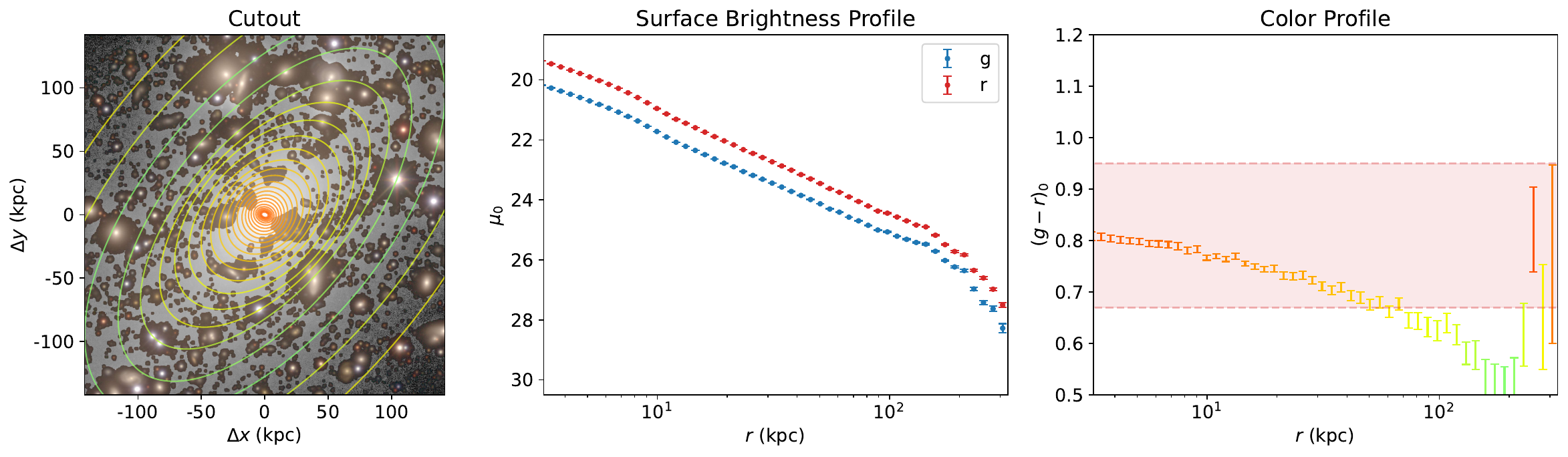}
    \includegraphics[width=0.98\textwidth]{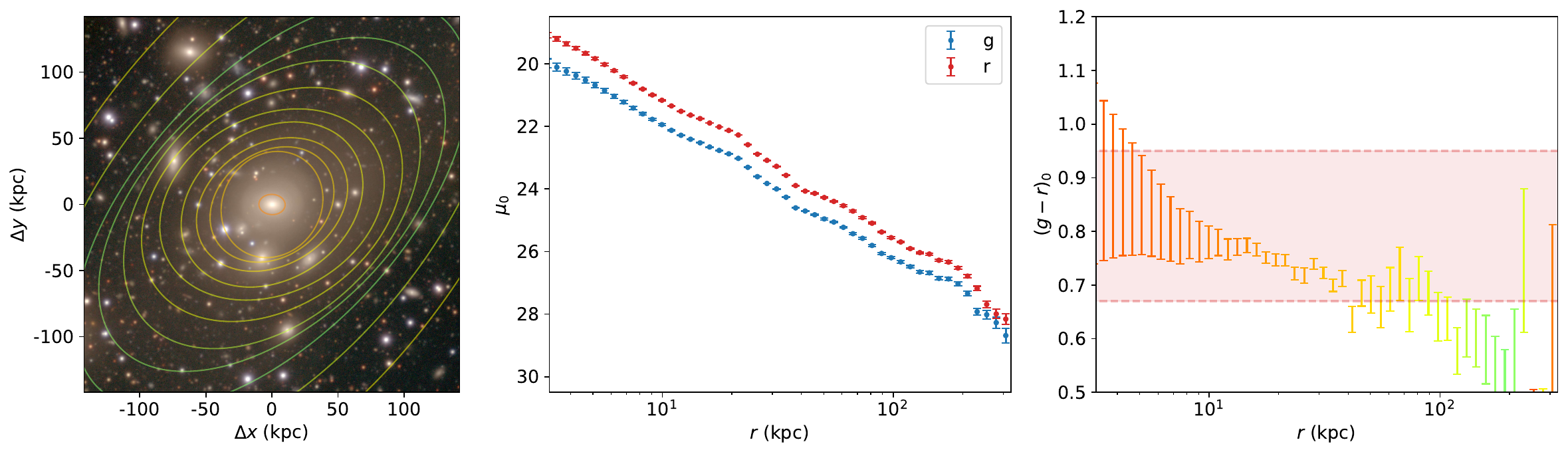}
    \includegraphics[width=0.98\textwidth]{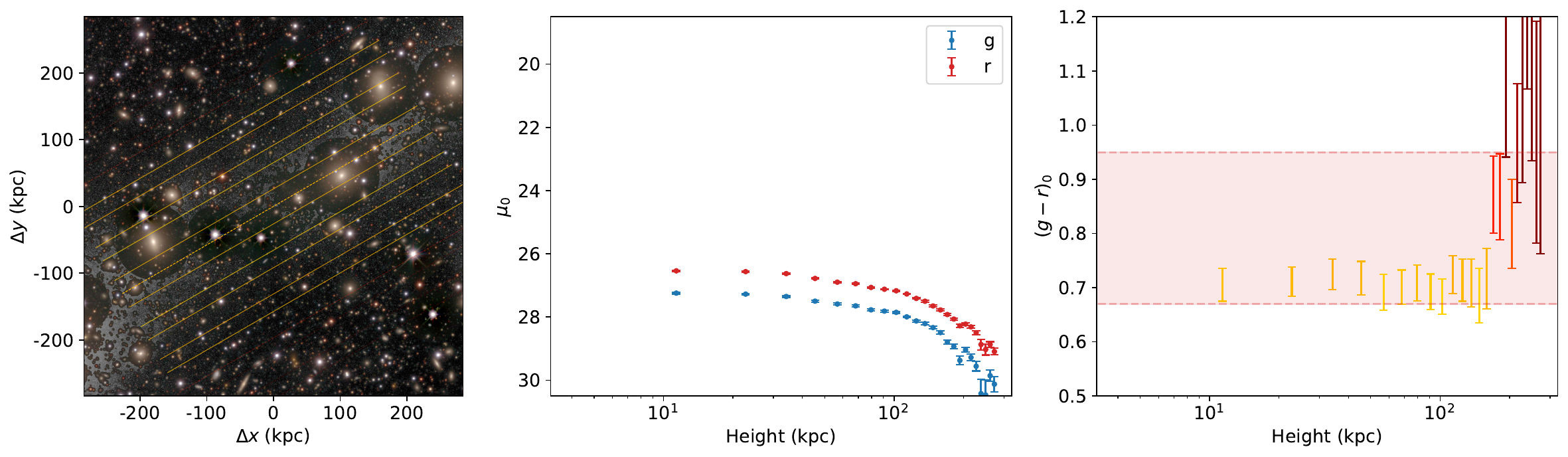}
    \caption{The calibrated surface brightness and color profiles for BCG1 (top), BCG2 (middle), and the bridge (bottom). The red shaded region in the color profiles covers the red sequence of A3667 \citep{fu_lovoccs_2024} The isophotes and annuli are drawn over $grz$ RGB cutouts and colored according to $(g-r)_0$. BCG1 is cropped to the core with the mask built in Section \ref{sec:dataproc} applied. We have rendered BCG2 without the mask and central isophotes to showcase the shells surrounding its core. The bridge has the modified LSP mask (detections, bad-pixels, stars, and intermediate galaxies) applied { and the $h=0$, the center, contour is rendered with a dashed line.}}
    \label{fig:bcg1}
\end{figure*}

To study the surface brightness and color profiles of the ICL, we used two approaches: fitting isophotes to galaxies and using rectangular bands to measure the bridge's profile. We fit isophotes to BCG1, BCG2, and LEDA 64218 using the \texttt{photutils.isophote} package \citep{bradley_photutils_2016} with background sources masked as outlined in Section \ref{sec:dataproc}. The fitting was carried out in two steps: an initial fit allowed the center of each isophote to vary and a final fit { fixed the center at the median center of the initial isophotes}. For the bridge, we built a series of rectangular bands which extend parallel to the line connecting BCG1 and BCG2. We assumed rectangular symmetry along this line and measured the bridge's profile by averaging the flux of the bands at the same distance above and below the line connecting the BCGs. Each profile has been transformed into the rest-frame by correcting for extinction \citep{schlegel_maps_1998}, applying a zero-point correction (Appendix \ref{app:depth}), a K-correction \citep{chilingarian_analytical_2010,chilingarian_universal_2012}, and correcting for cosmological surface brightness dimming.

To quantify the significance of the ICL signal, we define the signal to noise ratio per isophote (or band) as the ratio between the mean flux ($f$) and error on the mean flux ($\sigma_{f}$): $SN = \frac{f}{\sigma_{f}}$. The signal drops below our threshold for significance ($SN \lesssim 3$) at a distance of $\sim 150\arcsec$ ($160\text{ kpc}$) for the bridge, at a radius $\sim400\arcsec$ ($430\text{ kpc}$) for BCG1, and at a radius of $\sim200\arcsec$ ($200\text{ kpc}$) for BCG2 (Fig. \ref{fig:sn}). These end-points correspond to an observed surface brightness of $\sim 30 \text{mag}/\tarcsec^2$ (Fig. \ref{fig:lsb}), consistent with the limiting surface brightness of the observations.

\begin{figure*}[ht!]
    \centering
    \includegraphics[width=0.98\textwidth]{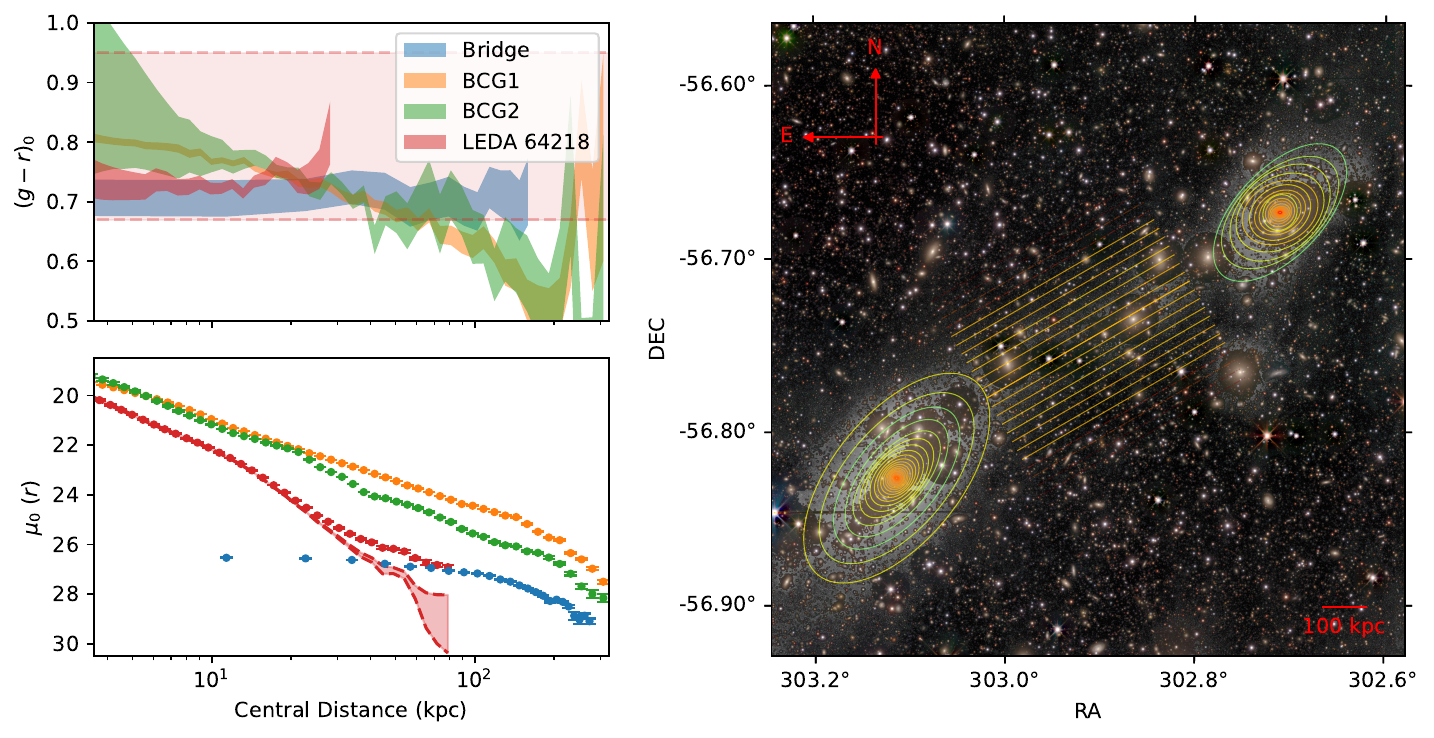}
    \caption{\textit{Left:} The rest-frame $g-r$ color and $r$-band surface brightness profiles; the dashed red curves in the bottom-left panel enclose the profile of LEDA 64218 minus the profile of the bridge. The color profiles have been truncated at smaller central distances for ease of viewing. \textit{Right:} A cropped cutout of the sky-corrected and star-subtracted image from Fig. \ref{fig:merged} with { statistically significant ($SN \gtrsim 3$)} isophotes and annuli rendered; these are colored per the $(g-r)_0$ color as in Fig. \ref{fig:bcg1}.}
    \label{fig:params}
\end{figure*}

% we might get some flack here since I'm not being quantitative in what it means to have a relaxed/disrupted profile
The profiles of BCG1, BCG2, and the bridge are presented in Fig. \ref{fig:bcg1}, and the profile of LEDA 64218 is presented in Fig. \ref{fig:params}. The surface brightness profiles surrounding BCG1 and BCG2 transition to the ICL at a radius of $\sim 100 \text{kpc}$, signaled by reaching a surface brightness dimmer than $\mu \sim 26 \text{ mag}/\tarcsec^2$ and a downturn in the profile. 
The profile of BCG1 appears consistent with previous studies \citep{montes_intracluster_2014,montes_intracluster_2018}, but the profile of BCG2 is disrupted due to the shells surrounding its core at $r\sim 20-50\text{ kpc}$ (Fig. \ref{fig:bcg1}).
%Comparing to previous studies \citep{montes_intracluster_2014,montes_intracluster_2018}, the profile of BCG1 is consistent with a relaxed-cluster while the profile of BCG2 is disrupted beyond the shells enclosing its core. 
The surface brightness profile of LEDA 64218 changes its behavior at $\sim 30 \text{ kpc}$ from the core and flattens to match the profile of the bridge at $\sim 70 \text{ kpc}$.

We compute the $g-r$ color profile of each object (Fig. \ref{fig:bcg1} and \ref{fig:params}). BCG1 and BCG2 have well-defined negative color gradients of similar slopes. BCG2, however, flattens at a distance of $\sim50-100\text{kpc}$ from the core. LEDA 64218 and the bridge are consistent with flat color-profiles.

%TODO draw the bridge and BCG1 in a single subplot

\section{Discussion and Conclusion} \label{sec:conc}

\subsection{The Buildup of ICL}

The steep negative color gradients surrounding BCG1 and BCG2 is consistent with ICL forming via gradual stripping of satellite galaxies \citep{contini_different_2018}. Unfortunately, with only a single color, the corresponding age and metallicity gradients cannot be studied in detail \citep[e.g., ][]{montes_buildup_2021}. We can partially infer their origin by comparing the color profile to the location of red sequence galaxies \citep{fu_lovoccs_2024}, which implies that stars forming the inner $\lesssim 60\text{ kpc}$ ($\lesssim 50\text{ kpc}$) of BCG1 (BCG2) originate from red sequence cluster members. 

The color profile in the outer region BCG2, from $\sim 50-100\text{ kpc}$, flattens, implying that the stars in this region have been mixed due to a recent merger which dissolves any potential age or metallicity gradients \citep{contini_origin_2021}. The bridge also has a flat color profile, implying that it has also been formed from a recent merger. On an initial inspection it is unclear if the stars composing the bridge originate from BCG2.

There are two likely scenarios under which the bridge could form: stars may be stripped from one BCG or galaxies between the BCGs are being stripped due to tidal forces across the cluster. These scenarios can be distinguished using the color of the bridge relative to other cluster members. The color of the bridge is consistent with the mixed region of BCG2 and does not significantly overlap with the color of LEDA 64218, moreover it is significantly larger than LEDA 64218 and other intermediate galaxies along the bridge, implying that the stars forming the bridge originate from BCG2. This is further supported when studying the profile of LEDA 64218 after the bridge has been subtracted. The profile is not disrupted, implying that stars in its outer regions are not accreting to form the bridge (Fig. \ref{fig:params}).

\subsection{The Formation of Abell 3667}

Based on this, we hypothesize that Abell 3667 is in the early stages of its merger, with BCG2 and BCG1 having undergone a nearly radial first pass. During this pass, BCG2 was disrupted by the merger which lead to the formation of shells around its core along with the extended plumes present in both BCGs \citep{valenzuela_stream_2024}. The merger has disrupted BCG2, mixing its stars and partially flattening its color profile. The resulting tidal forces have begun stripping stars from BCG2 that may be accreting onto BCG1, resulting in a flat color profile across the bridge.

Radio and X-ray observations of A3667 also support this hypothesis, with the current leading explanation being that A3667 is the product of an offset merger with a small impact parameter \citep{omiya_indications_2024}. The accretion of stars onto BCG1 can be confirmed if the color-profile of BCG1 increases to match the color of the bridge. We see an increase in BCG1's color at a distance of $\sim 200 \text{ kpc}$ from the core, but the ICL signal rapidly decays in this region so we do not definitively detect accretion.

\subsection{Conclusion}

{ Observations of equivalent depth, taken by LSST, are at least a decade away, leaving the coadds presented here as the deepest optical images of this field. In a future paper, we intend to carry out a full-spectrum analysis of this cluster which will use optical, X-ray, radio, and submillimeter observations (through the Sunyaev–Zel'dovich effect \citep{sunyaev_small-scale_1970,sunyaev_observations_1972,sunyaev_microwave_1980}) to determine the dynamical history of A3667. These will be compared with the latest simulations and used to study the initial conditions required to produce LSB features such as the bridge.}

Upcoming observations of A3667 among other local clusters by LSST are set to transform ICL studies by reaching and potentially exceeding the limiting surface brightnesses reached in this letter with at least six bands of photometry. This will place tight constraints on the age and metallicity gradients of the ICL, which could directly explain the origin and composition of features like A3667's bridge. In preparation for this, multiple teams are advancing background and star-subtraction algorithms, \citep{collaboration_lsst_2021,watkins_strategies_2024,bazkiaei_bright_2024,2023MNRAS.520.2484K}, exploring methods of modeling and subtracting galactic cirrus from LSST-like images \citep{liu_fuzzy_2025,smirnov_prospects_2023,roman_galactic_2020}, and considering new methods for analyzing the ICL \citep{brough_preparing_2024, jimenez-teja_unveiling_2018, ellien_dawis_2021}. Altogether, LSST observations will help constrain the stellar population of the ICL while also enabling detailed studies of the dynamical history of local clusters.

%The current era of wide-field multi-band surveys is exciting for the study of local clusters in general. MeerKAT observations of A3667 unveiled a radio halo coincident with the optical bridge and the peak X-ray luminosity of the cluster \citep{knowles_meerkat_2021,gasperin_meerkat_2022}. Teams such as the LoVoCCS and the Dark Energy Science Collaboration Clusters team are working to produce high-quality mass-maps of these observations with weak lensing \citep{the_lsst_dark_energy_science_collaboration_lsst_2018,aguena_clmm_2021}. The Nancy Grace Roman Space Telescope is will push local cluster studies further by providing wide-field observations with nearly diffraction-limited seeing and minimal sky background, which is ideal for lensing and LSB studies \citep{mantz_future_2019,montes_optimizing_2023}. This combination of future observations can be used to map the dark matter, stellar mass, and hot gas independently, allowing detailed studies of cluster assembly \citep{finner_hubble_2023,caretta_tracing_2023}.

%% Also note that the akcnowlodgment environment does not support long amounts of text. If you have a lot of people and institutions to acknowledge, do not use this command. Instead, create a new 
\section{Acknowledgments}
%\begin{acknowledgments}

{ We would like to thank the anonymous reviewer for their comments.} We would also like to thank the Observational Cosmology and Cluster Weak Lensing Group at Brown University and the Local Volume Complete Cluster Survey for their feedback on this draft. The authors would like to give a special thanks to Jacqueline McCleary, Michael Cooper, Brittany Torres, and Shenming Fu for their thorough comments on the earliest version of this draft.

AE acknowledges ongoing support from the NASA Rhode Island Space Grant Consortium.

ID and AE acknowledge support from the National Science Foundation (No. AST-2108287; Collaborative Research; LoVoCCS). ID and AE also acknowledge prior support from the U.S. Department of Energy, Office of Science under Award Number DE-SC-0010010.

MM acknowledges support from grant RYC2022-036949-I financed by the MICIU/AEI/10.13039/501100011033 and by ESF+, and program Unidad de Excelencia Mar\'{i}a de Maeztu CEX2020-001058-M.

This project used data obtained with the Dark Energy Camera (DECam), which was constructed by the Dark Energy Survey (DES) collaboration. Funding for the DES Projects has been provided by the DOE and NSF (USA), MISE (Spain), STFC (UK), HEFCE (UK), NCSA (UIUC), KICP (U. Chicago), CCAPP (Ohio State), MIFPA (Texas A\&M), CNPQ, FAPERJ, FINEP (Brazil), MINECO (Spain), DFG (Germany) and the Collaborating Institutions in the Dark Energy Survey, which are Argonne Lab, UC Santa Cruz, University of Cambridge, CIEMAT-Madrid, University of Chicago, University College London, DES-Brazil Consortium, University of Edinburgh, ETH Zürich, Fermilab, University of Illinois, ICE (IEEC-CSIC), IFAE Barcelona, Lawrence Berkeley Lab, LMU München and the associated Excellence Cluster Universe, University of Michigan, NSF NOIRLab, University of Nottingham, Ohio State University, OzDES Membership Consortium, University of Pennsylvania, University of Portsmouth, SLAC National Lab, Stanford University, University of Sussex, and Texas A\&M University.

Based on observations made at NSF Cerro Tololo Inter-American Observatory, NSF NOIRLab (Table \ref{tab:props}), which is managed by the Association of Universities for Research in Astronomy (AURA) under a cooperative agreement with the U.S. National Science Foundation.

This research draws upon DECam data as distributed by the Astro Data Archive at NSF NOIRLab.

This paper makes use of LSST Science Pipelines software developed by the \href{https://rubinobservatory.org/}{Vera C. Rubin Observatory}. We thank the Rubin Observatory for making their code available as free software at \href{https://pipelines.lsst.io}{https://pipelines.lsst.io}.

This research was conducted using computational resources and services at the Center for Computation and Visualization, Brown University.

%\end{acknowledgments}


\vspace{5mm}
\facilities{Blanco, Astro Data Archive}


%TODO add in photutils? numpy? scipy?
\software{LSST Science Pipelines \citep{bosch_hyper_2018,bosch_overview_2019},
          \texttt{astropy} \citep{collaboration_astropy_2022},
          \texttt{photutils} \citep{bradley_photutils_2016},
          \texttt{numpy} \citep{harris_array_2020},
          \texttt{scipy} \citep{virtanen_scipy_2020},
          \texttt{matplotlib} \citep{hunter_matplotlib_2007},
          Source Extractor \citep{bertin_sextractor_1996},
          }



\appendix

\section{Photometry \& Depth} \label{app:depth}

Our full procedure for photometric calibration is discussed in \citet{fu_lovoccs_2022}, but we briefly review the key details here. Following instrumental-signal removal, each detector was calibrated using reference stars matched with the appropriate filter in SkyMapper DR2 (with the exception of $u$-band, which we matched with SkyMapper's $v$-band) \citep{onken_skymapper_2019}. After stacking, we derived a per-band linear zero-point correction by transforming the DECam-magnitudes of reference-stars to SkyMapper with the appropriate color-terms. For $u$-band specifically, we used the stellar-locus to compute the zero-point correction \citep{william_high_slar_2009}.

The $5\sigma$ depth is presented in Table \ref{tab:depth}; based on the point-source depth, we have reached at least LSST year eight depth.

\begin{deluxetable}{cccccc}[htbp!]
\tablecaption{ The depth of our coadd based on point and extended sources selected within $1^{\circ}$ from the peak X-ray luminosity of A3667; estimated with the median magnitude of sources with $4.5 < SN < 5.5$. The magnitude of point and extended sources are measured using PSF and CModel magnitudes respectively \citep{abazajian_second_2004}. }\label{tab:depth}
\tablehead{ \colhead{Band} & \colhead{$5\sigma$ PSF} & \colhead{$5\sigma$ CModel} & \colhead{LSST Depth} }
\startdata
u & 25.2 & 24.9 & Y8 \\
g & 26.3 & 26.1 & Y8 \\
r & 26.3 & 26.0 & Y8 \\
i & 25.8 & 25.3 & Y8 \\
z & 25.2 & 24.8 & Y9 \\
\enddata
\end{deluxetable}

% Excluding the SN~10 depth, SN~5 is normally what is quoted
%u & 25.2 & 24.9 & 24.4 & 24.1 & Y8 \\
%g & 26.3 & 26.1 & 25.6 & 25.4 & Y8 \\
%r & 26.3 & 26.0 & 25.6 & 25.3 & Y8 \\
%i & 25.8 & 25.3 & 25.2 & 24.8 & Y8 \\
%z & 25.2 & 24.8 & 24.4 & 24.0 & Y9 \\

{

\section{Sky Correction}\label{app:sky}

\begin{figure*}[htbp!]
    \centering
    \includegraphics[width=0.96\textwidth]{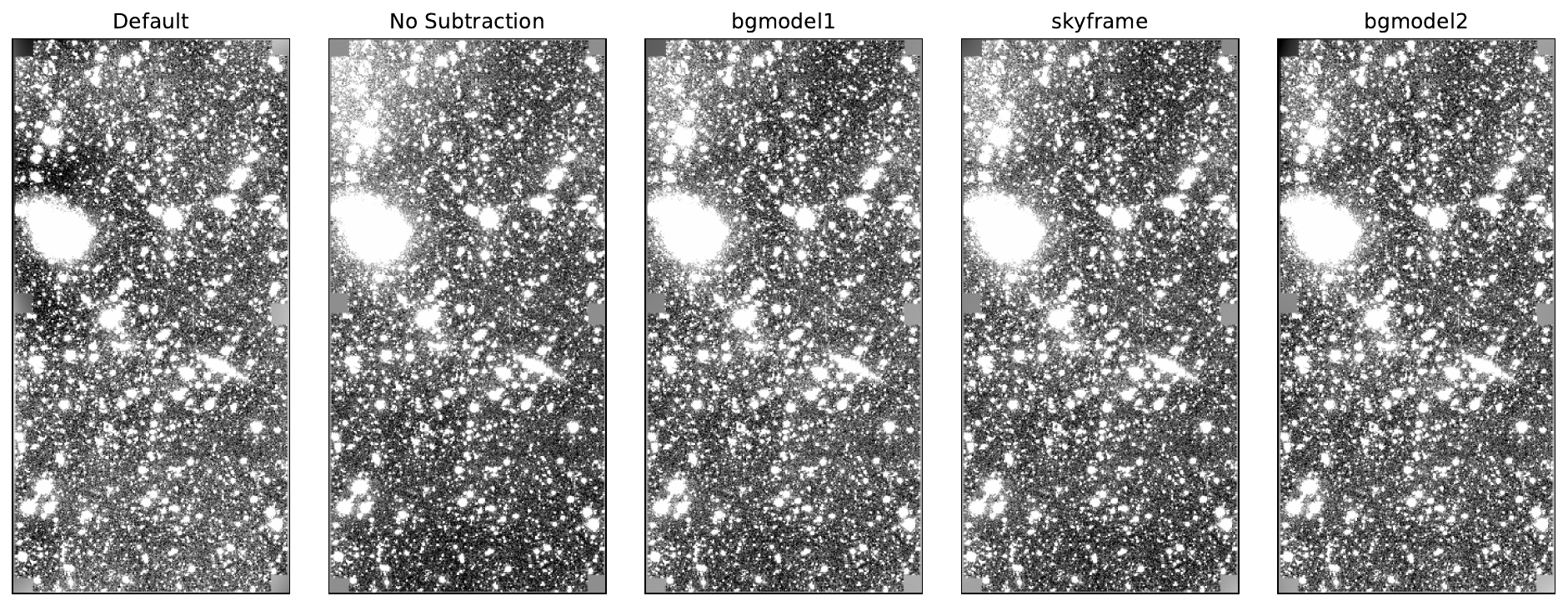}
    \caption{Detector 42 from DECam visit 1110874, binned into $8\text{ px} \times 8\text{ px}$ super-pixels with an aggressive stretch, shown during different steps of \texttt{skycorr}. From left to right: the default \texttt{calexp} produced by the LSP with an aggressive background subtraction, the same \texttt{calexp} with the background restored, \texttt{calexp} - \texttt{bg1}, \texttt{calexp} - \texttt{bg1} - \texttt{sky}, and \texttt{calexp} - \texttt{bg1} - \texttt{sky} - \texttt{bg2}. The default background model has a pronounced dark ring around a pair of merging galaxies. Without any model, there is a gradient across the detector. This is corrected by \texttt{bg1} and the skyframe, while preserving the LSB features such as the extended profile of the merging galaxy and the wings of a bright star outside the field of view (top-left). \texttt{bg2} does not noticeably flatten the background further and removes light from the LSB features.}
    \label{fig:detectors}
\end{figure*}

Studying ICL typically requires a specialized processing pipeline due to its sensitivity to systematics. This letter uses a modified version of the \texttt{skycorr} algorithm which is implemented in the LSP\footnote{\url{https://github.com/lsst/pipe_tasks/blob/26.0.0/python/lsst/pipe/tasks/skyCorrection.py}} \citep{aihara_second_2019,aihara_third_2022}. 

This algorithm depends on the creation of a skyframe for each band, which represents the mean response of DECam to the sky. For this letter, the skyframes were assembled using a large sample of exposures downloaded from the NOIRLab Astro Data Archive. A skyframe can vary depending on the observing conditions, so we built a selection of exposures whose conditions (e.g. the distribution airmasses, moon-phases, and moon-positions) were consistent with the conditions during the observation of A3667. We removed exposures separated by $<1^{\circ}$ and within $15^{\circ}$ of the galactic plane. This left $\sim 200$ exposures per-band, which were stacked to create each skyframe using the LSP's calibration-products pipeline (see Section \ref{sec:dataproc}).

The steps of \texttt{skycorr} are demonstrated in Fig. \ref{fig:detectors}, there are four key parts:

\begin{enumerate}
    \item The background of a calibrated exposure (calexp), which is stored separate from the exposure itself in the LSP, is restored. 
    \item Background pixels are binned into $4096\text{ px} \times 4096\text{ px}$ ($\sim 1000\arcsec \times 1000\arcsec$) super-pixels. These are smoothly interpolated \emph{across the focal plane} and subtracted from each detector (\texttt{bg1}). This is intended to remove large-scale gradients due to the moon and other sources of scattered light.
    \item Background pixels are binned into $32\text{ px} \times 32\text{ px}$ ($8\arcsec\times8\arcsec$) super-pixels. The skyframe (\texttt{sky}) is fitted to these super-pixels \emph{across the focal plane}, which is then interpolated and subtracted from each detector. This is intended to remove scattered light features which are fixed in the focal-plane, such as the "ghost-pupil" (see Fig. 4 of \citet{aihara_second_2019} or Fig. 4 of \citet{bernstein_instrumental_2017} for examples on HSC and DECam respectively).
    \item Background pixels are binned into $256\text{ px} \times 256\text{ px}$ super-pixels. These are smoothly interpolated \emph{across the focal plane} and subtracted from each (\texttt{bg2}). This is intended to remove small-scale features that are not modeled by the skyframe.
\end{enumerate}

These steps are iterated and, after each round of subtraction, source detection is ran again and the detection masks are updated for the next iteration (Fig. \ref{fig:masks}), the mask and background typically converges after two iterations. Fitting and subtraction are done using coordinates in the focal plane where the background is continuous. The results of this algorithm are showcased in Fig. \ref{fig:detectors}. The default background model used by the LSP aggressively over-subtracts, producing dark halos around large objects with extended features. The combination of \texttt{bg1} and the skyframe successfully flatten the background, without noticeably removing the light of extended features. However, \texttt{bg2} noticeably subtracts extended features larger than its super-pixels. Since the bridge extends well beyond the size of the super-pixels used for \texttt{bg2}, we disabled it for this letter.

\begin{figure}[htbp!]
    \centering
    \includegraphics[width=0.48\textwidth]{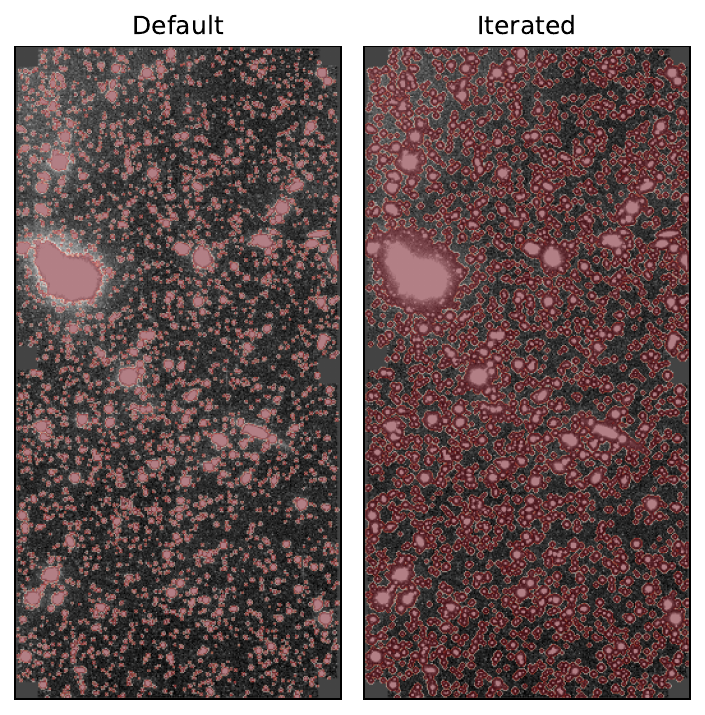}
    \caption{Detector 42 from DECam visit 1110874, shown with the default mask produced by the LSP and the final mask created by \texttt{skycorr} overlayed in red. The final mask is significantly more aggressive, but still fails to mask the most extended features, such as the wings of bright stars and the extended profile of the merging galaxies. This leaves them susceptible to over-subtraction by \texttt{bgmodel2}.}
    \label{fig:masks}
\end{figure}

}

\section{Proposal Information}\label{app:prop}

A table of the proposals and total exposure times used in this letter is provided in Table \ref{tab:props}. Over the past decade, A3667 has been observed with DECam by LoVoCCS and Weighing The Giants \citep{von_der_linden_weighing_2014}. Moreover, A3667 lies in both the Dark Energy Survey \citep{collaboration_dark_2016} and DeROSITA Survey \citep{salvato_erosita_2022} footprints. Combined with stray pointings from time-domain surveys and smaller projects, A3667 has accrued $\sim 32 \text{ hrs}$ exposure time in total.

\begin{deluxetable*}{ccccccc}[htbp!]
\tablecaption{ Datasets queried from the NOIRLab Astro Data Archive which were coadded for this letter along with the total exposure time per-band in seconds.}\label{tab:props}
\tablehead{ \colhead{Proposal ID} & \colhead{PI} & \colhead{u} & \colhead{g} & \colhead{r} & \colhead{i} & \colhead{z} }
\startdata
2012B-0001 & J. Frieman & - & 8280 & 6480 & 4680 & 6390 \\
2013A-0400 & J. Bloom & - & - & - & - & 1040 \\
2013A-9999 & A. Walker & - & - & 30 & - & - \\
2014A-0390 & J. Bloom & - & - & - & - & 480 \\
2014A-0415 & A. von der Linden & 3355 & 2355 & 4010 & 7850 & - \\
2014A-0624 & H. Jerjen & - & 720 & 720 & - & - \\
2014B-0244 & A. von der Linden & - & - & 3355 & 560 & 6980 \\
2015A-0618 & C. Lidman & 1000 & 500 & 500 & 500 & 1520 \\
2016A-0397 & A. von der Linden & 2606 & 4935 & 3435 & 3685 & 5760 \\
2018A-0242 & K. Bechtol & - & - & 1890 & 1260 & - \\
2019A-0308 & I. Dell'Antonio & 7360 & 3625 & 3740 & 2250 & - \\
2022A-597406 & A. Zenteno & - & 1800 & 1800 & 400 & - \\
2022A-975778 & K. Kelkar & - & 6000 & 3900 & - & - \\
2023A-585032 & A. Zenteno & 810 & - & - & - & - \\
2023B-646244 & A. Chiti & - & 90 & - & - & - \\
\hline
 & {\bf Total} & 15131 & 28305 & 29560 & 21185 & 22170 \\
\enddata
\end{deluxetable*}


\bibliography{zotero_5-28-25_nodupe}{}
\bibliographystyle{aasjournal}



\end{document}